\documentclass[%
 pra,reprint,
 showpacs,
 amsmath,amssymb,
 aps,
]{revtex4-1}

\usepackage{graphics,graphicx}
\graphicspath{{./}}
\usepackage{epsf}
\usepackage{subfigure}
\usepackage[%
  colorlinks=true,
  urlcolor=blue,
  linkcolor=blue,
  citecolor=blue
]{hyperref}
\usepackage{dcolumn}
\usepackage{bm}
\usepackage{color}
\usepackage{textpos}

\begin{document}

\preprint{APS/123-QED}

\title{Static kinks in chains of interacting atoms}

\author{Haggai Landa}\email{haggaila@gmail.com}\thanks{Current address: IBM Research - Haifa, Haifa University Campus, Mount Carmel, Haifa 31905, Israel}
\affiliation{Institut de Physique Th\'{e}orique, Universit\'{e} Paris-Saclay, CEA, CNRS, 91191 Gif-sur-Yvette, France}
\author{Cecilia Cormick}
\affiliation{IFEG, CONICET and Universidad Nacional de C\'ordoba, 
X5016LAE C\'{o}rdoba, Argentina}
\author{Giovanna Morigi}
\affiliation{Department of Physics, Universit\"at des Saarlandes, D-66123 Saarbr\"ucken, Germany}

\date{\today}%

\begin{abstract}
We theoretically analyse the equation of topological solitons in a chain of particles interacting via a repulsive power-law potential and 
confined by a periodic lattice. Starting from the discrete model, we perform a gradient expansion and obtain the kink 
equation in the continuum limit for a power-law exponent $n\ge 1$. The power-law interaction modifies the 
sine-Gordon equation, giving rise to a rescaling of the coefficient multiplying the second derivative (the kink width) and to an additional integral term. We argue that the integral term does not affect the local properties of the kink, but it governs the behaviour at the asymptotics. The kink behaviour at the center is dominated by a sine-Gordon equation and its width tends to increase with the power law exponent. When the interaction is the Coulomb repulsion, in particular, the kink width depends logarithmically on the chain size. We define an appropriate thermodynamic limit and compare our 
results with existing studies performed for infinite chains. Our formalism allows one to 
systematically take into account the finite-size effects and also slowly varying external 
potentials, such as for instance the curvature in an ion trap. 
 \end{abstract}


\maketitle

\section{Introduction}

The Frenkel-Kontorova model reproduces in one dimension the essential features of stick-slip motion 
between two surfaces \cite{Frenkel-Kontorova,Bak:RPP,Tosatti:RMP}. The ground state is expected to 
describe the structure of a one-dimensional crystal monolayer growing on top of a substrate 
crystal. In one dimension the elastic crystal is modelled by a periodic chain of classical 
particles with uniform equilibrium distance $a$, which interact with a sinusoidal potential 
with periodicity $b$ \cite{Pokrovski:Book,Kivshar:Book}. Frustration emerges from the competition 
between the two characteristic lengths: Depending on the mismatch between $a$ 
and $b$ and on the strength of the substrate potential, a continuous transition occurs between a 
structure with the substrate's period (commensurate) and an incommensurate structure \cite{Talapov}. 
The transition is characterized by proliferation of kinks, namely, of local distributions of excess 
particles (or holes) in the substrate potential. When the interactions of the elastic crystal are 
nearest-neighbour, in the long-wavelength limit the dynamics of a single kink is governed by the 
integrable sine-Gordon equation \cite{Merwe,Talapov,Tabor}.


The experimental realizations of crystals of interacting atoms, such as ions \cite{Dubin:RMP}, 
dipolar gases \cite{Ferlaino}, and Rydberg excitons \cite{Rydberg:Review}, offer unique platforms 
for analysing the Frenkel-Kontorova model dynamics \cite{Garcia-Mata,Pruttivarasin,Vanossi}. The 
substrate potential can be realised by means of optical lattices 
\cite{Linnet:2012,Enderlein:2013,Ferlaino,Cetina:2013} or of a second atomic crystal 
\cite{Drewsen,Mehlstaeubler}. Periodic boundary conditions can be implemented in ring traps 
\cite{Haeffner}. Kinks and dislocations can be imaged 
\cite{Pyka,Ulm,Mielenz,Monroe,Atom:microscope,Rydberg:Arimondo} and spectroscopically analysed 
\cite{Brox}. Differing from textbook models, the particles' interaction is a power-law potential, 
whose exponent could be engineered by means of lasers \cite{Porras:RPP}. 
Nano-friction have been experimentally investigated in small ion chains in periodic 
potentials \cite{Vuletic:1,Vuletic:2,Vuletic:3}.

The study of kinks and of nanofriction in these systems  requires one to analyse the effects of the tails of the interactions
on the sine-Gordon equation. Specifically, in one dimension the energy is non additive in Coulomb systems \cite{Kac}. Yet, long-range effects are marginal: the exponent of the Coulomb interaction formally separates two regimes, such that for slower power-law decays the dynamical equations are characterized by fractional spatial derivatives, while for faster decays the spatial derivatives are of integer order \cite{Zavlasky,Staffilani}. The effect of long-range interactions on the commensurate-incommensurate transition have been discussed \cite{Bak:PRL,Pokrovsky:LR},
the kink solutions in a periodic potentials have been analysed numerically for the long-range Kac-Baker interactions  \cite{Kac-Baker}
and for dipolar and Coulomb potentials \cite{Braun:1990}. Analytic studies of the kink solutions have been performed in the thermodynamic limit \cite{Kac-Baker,Pokrovsky:LR,Braun:1990}. 

The aim of the present work is to review the analytical derivation of the kink equation for power-law interacting potentials by means of a gradient expansion, which is implemented following the lines of the  study of Ref. \cite{Morigi:2004,Morigi:2004:PRL}. This derivation allows one to determine the local properties 
of the kink as a function of the interaction range for integer exponents when this decays with the distance as the Coulomb repulsion or faster. Moreover, it allows one to determine its asymptotic behaviour, as well as to 
systematically take into account the finite-size effects, thus setting the basis of a study where 
these effects can be included in a perturbative fashion. 

This manuscript is organised as follows. In Sec. \ref{Sec:2} we introduce the Lagrangian of an 
atomic chain in a periodic potential, where the atoms interact via a power-law potential. In Sec. 
\ref{Sec:3} we consider the long-wavelength limit and derive the equation for the static kink. We 
discuss separately the case of Coulomb interactions. We analyse then the thermodynamic limit and 
compare our results with the ones of Ref. \cite{Braun:1990}. The conclusions are drawn in Sec. 
\ref{Sec:4}.

\section{An atomic chain in a periodic substrate potential}
\label{Sec:2}

A chain of $N$ interacting atoms with mass $m$ is confined in a finite volume and is parallel to the $x$ axis. Their atomic positions and canonically-conjugated momenta are $x_j$ and $p_j$, with $j=-N/2,N/2+1\ldots,N/2-1$ and
$x_j<x_{j+1}$. Their Lagrangian $\mathcal L$ 
reads\begin{equation}
\mathcal L=\sum_{i=1}^N\frac{m\dot x_j^2}{2}-V_{\rm pot}\,,
\end{equation}
where $V_{\rm pot}$ is the potential energy and is thus the sum of a periodic substrate potential and of 
the harmonic interaction between pairs of particles, $V_{\rm pot}=V_{\rm opt}+V_{n}$.
The periodic substrate potential $V_{\rm opt}$ is a sinusoidal lattice with periodicity $b$ and amplitude 
$V_0$:
\begin{equation}
V_{\rm opt}=\sum_{i=1}^NV_0\left[ 1-\cos\left(\frac{2\pi x_i}{b}\right)\right]\,,
\end{equation}
The atomic interaction $V_n$ couples atoms at distance $r a$ with strength scaling as $1/r^{n+2}$:
\begin{equation}
V_{n}=\frac{1}{2}\sum_{i}\sum_{r>0} \frac{K_n}{r^{n+2}}( x_{i+r}-x_i-ra)^2\,,
\end{equation}
Here, $n$ an integer number, $n\ge 1$, and $K_n$ is the spring constant between nearest neighbour. The interaction term vanishes when the atoms are at the equilibrium positions  $x_j^{(0)}=ja$.

We note that interactions of this form are obtained by expanding the interaction potential till second order in the displacements about the equilibrium positions of the interaction forces, and discarding anharmonicities. Let the interaction between two particles at distance $x$ be given by 
\begin{equation}
W_{\rm int}(x)=W_n/x^n\,,
\end{equation} 
where $W_n$ is a constant which depends on $n$. Then, the spring constant takes the form
\begin{equation}
\label{K:n}
K_n=\partial_x^2W_{\rm int}(x)|_{x=a}=\frac{n(n+1)W_n}{a^{n+2}}\,.
\end{equation}
The textbook limit, where the particles interact via nearest-neighbour interactions, is recovered 
by letting $n\to\infty$. In this case $W_n$ shall be appropriately rescaled in order to warrant that the spring constant $K_\infty$ remains finite, with
 $K_\infty=\lim_{n\to\infty}K_n$.  In the following we discuss the cases of $n>1$ (with 
dipolar and van der Waals interactions corresponding to the particular values $n=3$ and $n=6$ 
respectively), as well as the Coulomb interaction $n=1$. For Coulomb interaction between particles of charge $q$, the spring constant is 
$K=2q^2/(4\pi\varepsilon_0a^3)$, with $q$ the charge of the particles and $\varepsilon_0$ the 
vacuum's permittivity \cite{Morigi:2004}. 

Here and in what follows we assume motion along one axis and open boundary conditions. This can be 
realised by means of anisotropic traps and sufficiently cold atoms, where one can assume that 
the motion in the transverse direction is frozen out. Typically the trap curvature gives rise to 
inhomogeneity in the equilibrium particle distribution, which leads to a position-dependent spring 
constant $K_n$. Below we assume that the atoms are uniformly distributed. We note, however, that 
this formalism can systematically include the trap inhomogeneity, as shown in Ref. 
\cite{Morigi:2004}, as long as the trap curvature can be treated in the continuum limit \cite{Dubin:1997}.

\subsection{Equilibrium configuration in the discrete chain}

The equilibrium configurations of the discrete chain are solutions of the equation of motion 
\begin{eqnarray}
\label{eq:kink:discrete}
m\ddot x_j&=&\sum_{r>0}\frac{K_n}{r^{n+2}}\left[x_{j+r}-x_j-(x_j-x_{j-r})\right]\\
& &-\frac{2\pi}{b}V_0\sin\left(\frac{2\pi x_i}{b}\right)\,,\nonumber
\end{eqnarray}
which satisfy $m\ddot x_j=0$ for all $j$. For $V_0=0$, namely, in the 
absence of the substrate potential, the ground state is the uniform chain 
with equilibrium positions
$$x_j^{(0)}=ja\,.$$
For $V_0\neq 0$ the equilibrium configuration $\{\bar{x}_j^{(0)}\}$ reads
$$\bar{x}_j^{(0)}= x_j^{(0)}+\bar u_j=ja +\bar u_j\,,$$
where $\bar u_j$ are static displacements that solve the set of equations: 
\begin{align}
\frac{2\pi V_0}{b} \sin\left[2\pi(x_i^{(0)}+\bar u_i)/b)\right]
=\sum_{r\neq 0} \frac{K_n}{|r|^{n+2}}(\bar u_{i+r}-\bar u_i)\,.
\label{eq:eom0}
\end{align}
A static kink describes particles displacements which are localized in a region of the chain, such 
that the chain is uniform at the edges. The solution interpolates between the two boundary values 
$u_j\to 0$ for $j\to -N/2$ and $u_j\to b$ for $j\to +N/2$ and describes a topological soliton. The antikink is the topological soliton of opposite charge and interpolates between 
$u_j\to 0$ for $j\to -N/2$ and $u_j\to -b$ for $j\to +N/2$. 
 
It is convenient to introduce the phase variable $\theta_j$ of particle $j$, which is a dimensionless scalar and is defined as
\begin{equation}
 \label{eq:thetadef0}
\theta_j=2\pi (\bar x_j^{(0)}-\ell bj)/b\,,
\end{equation}
or, alternatively:
\begin{equation}
\bar x_j^{(0)}=j\ell b+\frac{b}{2\pi}\theta_j\,.
 \label{eq:thetadef}
 \end{equation}
The phase function $\theta_j$ gives the shift of the ion $j$ from 
the commensurate configuration, such that for every $2 \pi$ change in the phase the respective ion 
position is shifted by one period $b$ of the periodic potential. The equation of the phase function 
is then given by 
\begin{eqnarray}
\label{eq:eomdiscrete0}
&-\sum_{r\neq 0} 
\frac{(\theta_{i+r}-\theta_i-(\delta_{i+r}-\delta_i))}{|r|^{n+2}}+m_K^2\sin(\theta_i)=0\,,
\end{eqnarray}
where 
\begin{equation}
\delta_j=2\pi(x_j^{(0)}-\ell bj)/b\,,
\label{deltaj}
\end{equation}
and represents the mismatch between the former equilibrium positions of the crystal and the nodes 
of the periodic potential. The equation depends now on the power-law exponent,  on 
the mismatches $\delta_j$, and more specifically, on the ratio $a/b$, and on the dimensionless 
ratio
\begin{equation}
\label{mK}
m_K=2\pi\sqrt{\frac{V_0}{Kb^2}}\,,
\end{equation}
which scales the weight of the interactions with the respect to the localizing potential. We now conveniently rewrite 
the mismatch by introducing the ratio $\ell=\lfloor a/b \rceil$, with $\lfloor  x \rceil$ the 
nearest integer to $x$. The configuration is commensurate when the ratio $a/b=\ell$, then the 
ground-state equilibrium positions are $x_j^{(0)}=ja=\ell jb$. When instead $\ell\neq a/b$ the 
displacement is generally $\bar u_j\neq 0$. For 
$x_j^{(0)}=ja$ in Eq. \eqref{deltaj}, one obtains $\delta_j=j\delta$ with 
\begin{equation}
\delta=2\pi (a-\ell b)/b\,,
\end{equation}
which vanishes when both length scales are commensurate with each other. In general, $\delta$ is a
periodic function of the ratio $a/b$, in fact $\delta(a/b+p)=\delta(a/b)$, $p\in \mathbb{Z}$, and 
its value ranges in the interval $-\pi <\delta<\pi$. We use this expression and rewrite Eq. 
\eqref{eq:eomdiscrete0} in terms of the phase function:
\begin{eqnarray}
\label{eom:discrete}
-\sum_{r\neq 0} \frac{(\theta_{i+r}-\theta_i-r\delta)}{|r|^{n+2}}+m_K^2\sin(\theta_i)=0\,.
\end{eqnarray}
In the rest of this manuscript we will consider structures whose ground state is commensurate, and 
thus take $\delta=0$ in Eq. \eqref{eom:discrete}. We analyse the equation of motion of single 
topological solitons in the long-wavelength limit when the periodic substrate potential is a small 
perturbation to the elastic crystal, $m_K\ll1$.


\section{Equation of the static kink in the long-wavelength limit}
\label{Sec:3}

In this section we derive the continuum limit of Eq.~\eqref{eq:eomdiscrete0} extending the 
procedure of Ref. \cite{Morigi:2004} to kink's equation. In our treatment we keep $n$ finite and consider the 
commensurate case, $a=\ell b$, when the mismatch $\delta$ vanishes. For completeness, we first shortly 
review the derivation of the sine-Gordon equation in the case of nearest-neighbour interactions and briefly discuss
the properties of its exact solution.

\subsection{Nearest-neighbour interactions}

The sine-Gordon equation of the Frenkel-Kontorova model is found in our model by considering the 
limit $n\to \infty$ in Eq. \eqref{eom:discrete}, thus keeping only the nearest neighbour terms 
($r=1$) and taking $K_\infty=\lim_{n\to\infty} K_n$ to be constant. The long-wavelength limit is 
determined by first taking the Fourier transform of the phase function, 
$\theta_j\to\tilde{\theta}(k)=\sum_j{\rm e}^{{\rm i}kx_j}\theta_j/\sqrt{N}$ and expanding the equation of motion for small $k$ \cite{Zavlasky}. In $k$ space the 
interaction term takes the form $-2K_{\infty} (1-\cos(ka))\tilde\theta(k)\sim -k^2a^2 K_{\infty} 
\tilde\theta(k)$. Going back to position space, this term is cast in terms of a second-order 
derivative of the phase function. Then, Eq. \eqref{eom:discrete} takes the form of a 
sine-Gordon equation \cite{Tabor}:
\begin{eqnarray}
\label{eq:SG:1}
-a^2\frac{\partial^2}{\partial x^2} \theta(x) +m_K^2\sin(\theta(x))=0\,.
\end{eqnarray}
One can rewrite the equation as 
\begin{eqnarray}
\label{eq:SG:2}
-d^2\frac{\partial^2}{\partial x^2} \theta(x) +\sin(\theta(x))=0\,,
\end{eqnarray}
where $d$ is the width of the kink \cite{Braun:1990},
\begin{equation}
\label{eq:d}
d=a\,\sqrt{\frac{K_\infty}{V_0}\left(\frac{b^2}{4\pi^2}\right)}=\frac{a}{m_K}\,.
\end{equation}
Equation \eqref{eq:SG:2} admits the solution \cite{Tabor}
\begin{equation}
\theta(x)=4 \tan^{-1}[e^{\sigma(x-x_0)/d}]\,,
\label{eq:thetaSG}
\end{equation}
with $\sigma=\pm 1$ the topological charge and $x_0$ the position of the kink. Specifically, for $\sigma=1$, the solution is a single kink with $\theta(x)\to 2\pi$ for $x\to \infty$, and 
$\theta(x)\to 0$ for $x\to -\infty$. The antikink corresponds to $\sigma=-1$ and is thus the mirror reflection of the kink at $x_0$. We note that the continuum limit is consistent when the width 
of the kink is much larger than the interparticle distance, $d\gg a$, and the potential is a 
small perturbation to the elastic crystal. 

The constant $m_K$ scales the mass $m_{\rm SG}$ of the sine-Gordon kink. We recall that the kink's mass $m_{\rm kink}$ is defined as 
$m_{\rm kink}=m\sum_j(\partial \bar u_j/\partial x_0)^2$. In the continuum limit and using Eq. \eqref{eq:thetaSG}, one obtains \cite{Braun:1990}
\begin{equation}
m_{\rm SG}=\frac{8m_K}{a}\,m\left(\frac{b}{2\pi}\right)^2 \label{m:SG}\,.
\end{equation}

Anharmonicities of the potential give rise to higher-order derivatives and thus to an asymmetry between kink and antikink \cite{Willis,Braun:1990}. In this treatment we discard these terms, restricting the expansion of the interaction potential to the harmonic terms. In this limit the kink and antikink in the continuous limit just differ because of the topological charge $\sigma$. We further note that, by performing the continuum limit, we discarded higher order terms in the gradient expansion. These terms account for discreteness effects and give rise to an effective kink narrowing \cite{Braun:1990}.

\subsection{Power-law interactions}

We now perform a gradient expansion for power-law interactions. This is done by means 
of a manipulation of Eq. \eqref{eom:discrete} which is equivalent to a Taylor expansion for low momenta 
in Fourier space, and consists in singling out the terms contributing to the second 
derivative.  The equation of motion for the static kink then becomes an 
integro-differential equation. In the following we will assume $m_K\ll 1$, unless otherwise specified. 

In order to study the long-wavelength limit we use the prescription $2i/N\to \xi$ and $\theta_j\to 
\theta(\xi)$. For $N\gg 1$, then $\xi$ can be treated as a continuous variable defined in the 
interval $[-1/2,+1/2]$. With this prescription Eq. \eqref{eom:discrete} takes the form:
\begin{equation}
-\frac{(d_n/a)^2}{N^{n+1}}\mathcal I_n+ \sin[\theta(\xi)]=0\,,
\label{eq:eom-continuum}
\end{equation}
where $\mathcal I_n$ is dimensionless:
\begin{align}
\mathcal I_n=&\int_{\bar 
a}^{1/2}d\xi'\frac{1}{\xi'^{n+2}}[\theta(\xi+\xi')-2\theta(\xi)+\theta(\xi-\xi')]\,,
\label{eq:inteq}
\end{align}
and the discrete nature of the chain at atomic distances enters through the (high-frequency) cutoff
$$\bar a=1/N\,.$$
In Eq. \eqref{eq:eom-continuum} we have also introduced the characteristic length
\begin{equation}
\label{d:n}
d_n=a\,\sqrt{\frac{K_n}{V_0}\left(\frac{b}{2\pi}\right)^2}\,,
\end{equation}
which would correspond to the kink's width when simply truncating the sum in Eq. 
\eqref{eom:discrete} till the nearest-neighbours. We note that in this formalism the chain length is a low-frequency cutoff, and in particular the limit 
$N\gg 1$ is equivalent to small $k$ in Fourier space. We will first keep $N$ constant, thus 
consider a finite chain, and take the thermodynamic limit only after performing the gradient expansion.

We now integrate  Eq. \eqref{eq:inteq} by parts applying the procedure as in Ref. 
\cite{Morigi:2004} and rewrite it as the sum of three terms \cite{Morigi:2004}: 
\begin{equation}\mathcal I_n=  I_{\rm edge}+I_{\bar a}+I_n\,,
\label{eq:I}
\end{equation}
where the first term on the RHS contains the contributions of the chain's edges:
\begin{align}
I_{\rm 
edge}=-&\frac{1}{(1+n)}\frac{[\theta(\xi+1/2)+\theta(\xi-1/2)-2\theta(\xi)]}{(1/2)^{n+1}}\notag \\ 
&-\frac{1}{2 n(1+n)}\frac{[\theta'(\xi+1/2)-\theta'(\xi-1/2)]}{(1/2)^{n+1}}
\,,
\label{I:E:0}
\end{align}
and $\theta'(\xi)\equiv\partial \theta(\xi)/\partial\xi$. The second integral describes the 
contribution of the nearest-neighbour terms:
\begin{align}
 I_{\bar a}=&\frac{1}{(1+n)}\frac{[\theta(\xi+\bar a)-2\theta(\xi)+\theta(\xi-\bar a)]}{\bar 
a^{n+1}}\notag \\ 
&+\frac{\bar a}{n(1+n)}\frac{[\theta'(\xi+\bar a)-\theta'(\xi - \bar a)]}{\bar a^{n+1}}\,. 
\label{I:bara:0}
\end{align}

By making a Taylor expansion about $\xi$, $I_{\bar a}$ can be cast 
into the form :
\begin{align}
\label{I:bara:0}
I_{\bar a}=&\frac{\bar a^2}{\bar 
a^{n+1}}\left(\frac{1}{1+n}+\frac{2}{n(1+n)}\right)\left(\theta''(\xi)+{\rm O}(\bar 
a^2\theta^{(4)}(\xi))\right)\,,
\end{align}
with $\theta''(\xi)\equiv\partial^2 \theta(\xi)/\partial\xi^2$ and $\theta^{j}(\xi)\equiv\partial^j 
\theta(\xi)/\partial\xi^j$.
Finally, the term $I_n$ reads:
\begin{eqnarray}
\label{I:alpha}
I_{n}&=&\frac{1}{n(n+1)}\int_{\bar 
a}^{1/2}d\xi'\frac{\theta''(\xi+\xi')+\theta''(\xi-\xi')}{\xi'^{n}}\,,
\end{eqnarray}
and still contains second derivatives of the phase function. We note that it can be also rewritten 
in the form:
\begin{eqnarray}
\label{I:alpha:Braun}
I_{n}&=&\frac{1}{n(n+1)}\frac{\partial}{\partial \xi}\int_{\bar 
a}^{1/2}d\xi'\frac{\theta'(\xi+\xi')+\theta'(\xi-\xi')}{\xi'^{n}}\,, 
\end{eqnarray}
so that the equation for the static kink is the integro-differential equation:
\begin{eqnarray}
\label{Kink:Braun}
&-&d_n^2\frac{n+2}{n(n+1)}\frac{\partial^2 \theta}{\partial x^2}+\sin\theta\\
& &=\frac{1}{n(n+1)}\frac{d_n^2}{a^2}\frac{1}{N^{n+1}}\frac{\partial}{\partial \xi}\int_{\bar 
a}^{1/2}d\xi'\frac{\theta'(\xi+\xi')+\theta'(\xi-\xi')}{\xi'^{n}}\nonumber\,.
\end{eqnarray}
Here, the contribution due to $I_{\rm edge}$ has been discarded, and we have used $N\bar a=1$ and 
$x=Na\xi$. Moreover, we have discarded higher order derivatives $\theta^{(2j)}(a)$, which account for discreteness effects. 
If we would take now the thermodynamic limit, and thus let $N\to\infty$, then this expression would coincide with the one reported in Ref. \cite{Braun:1990}, apart for the definition of the scaling coefficients and for higher order local derivatives. We note, however, that the integral term in Eq. \eqref{Kink:Braun} still contains terms which can be of the same order as $k^2a^2$. In order to single them out, we perform a further step 
of the partial integration. We identify two qualitatively different cases, the case $n>1$ and the 
Coulomb case $n=1$, which we discuss individually below. 

\subsubsection{Power-law interactions with $n>1$}

Performing partial integration of $I_n$ for $n>1$ we obtain 
\begin{eqnarray}
I_{n}&=&-\frac{1}{n(n^2-1)}\frac{\theta''(\xi+\xi')+\theta''(\xi-\xi')}{\xi'^{n-1}}\Bigl|_{\bar 
a}^{1/2}+I_n'\,,
\end{eqnarray}
and
\begin{eqnarray}
\label{I:n}
I_{n}'&=&\frac{1}{n(n^2-1)}\int_{\bar 
a}^{1/2}d\xi'\frac{\theta^{(3)}(\xi+\xi')-\theta^{(3)}(\xi-\xi')}{\xi'^{n-1}}\,,
\end{eqnarray}
We now collect the edge contributions and verify that they scale like $1/N^{n-1}$, thus we neglect 
them under the reasonable assumption that the kink derivatives vanish at the edges. Using that 
$N\bar a=1$ and going back to dimensional coordinates ($x=Na\xi$), we obtain the 
integro-differential equation for a static kink in a chain of atoms interacting via power-law 
interactions:
\begin{eqnarray}
\label{Kink:n}
&-&\frac{d_n^2}{n-1}\frac{\partial^2}{\partial x^2}\theta(x)+\sin\theta(x)=\\
& &
=\frac{d_n^2}{n-1}\frac{a^{n-1}}{n(n+1)}\int_{a}^{L/2}dx'\frac{\theta^{(3)}(x+x')-\theta^{(3)}
(x-x')}{x'^{n-1}}\,,\nonumber
\end{eqnarray}
where $L=Na$ is the chain's length, and now the third-order derivative is taken with respect to the 
dimensional coordinates.

\subsubsection{Coulomb interactions}

Fore $n=1$ partial integration of Eq. \eqref{I:alpha} leads to the expression
\begin{eqnarray}
I_{n=1}&=&\frac{1}{2}(\theta''(\xi+\xi')+\theta''(\xi-\xi'))\log \xi'\Bigl|_{\bar a}^{1/2}+I_1'\,,
\end{eqnarray}
where
\begin{eqnarray}
\label{I:1}
I_{1}'&=&-\frac{1}{2}\int_{\bar a}^{1/2}d\xi'(\theta^{(3)}(\xi+\xi')-\theta^{(3)}(\xi-\xi'))\log 
\xi'\,.
\end{eqnarray}
Neglecting the contributions from the edges we obtain the equation for a static kink in a 
sufficiently long chain of single-component charges:
\begin{eqnarray}
\label{Kink:LR}
&-&d_1^2\left(\frac{3}{2}+\log N\right)\frac{\partial^2}{\partial 
x^2}\theta+\sin\theta(x)\\
&&=-\frac{d_1^2}{2}\int_{a}^{L/2}dx'(\theta^{(3)}(x+x')-\theta^{(3)}(x-x'))\log (x'/L)\,.\nonumber
\end{eqnarray}

\subsection{Discussion}

\begin{widetext}

\begin{figure}[t]
\subfigure[\,$N=101$]{
\includegraphics[trim = 0cm 0cm 0.cm 0cm, clip,width=0.32\textwidth]{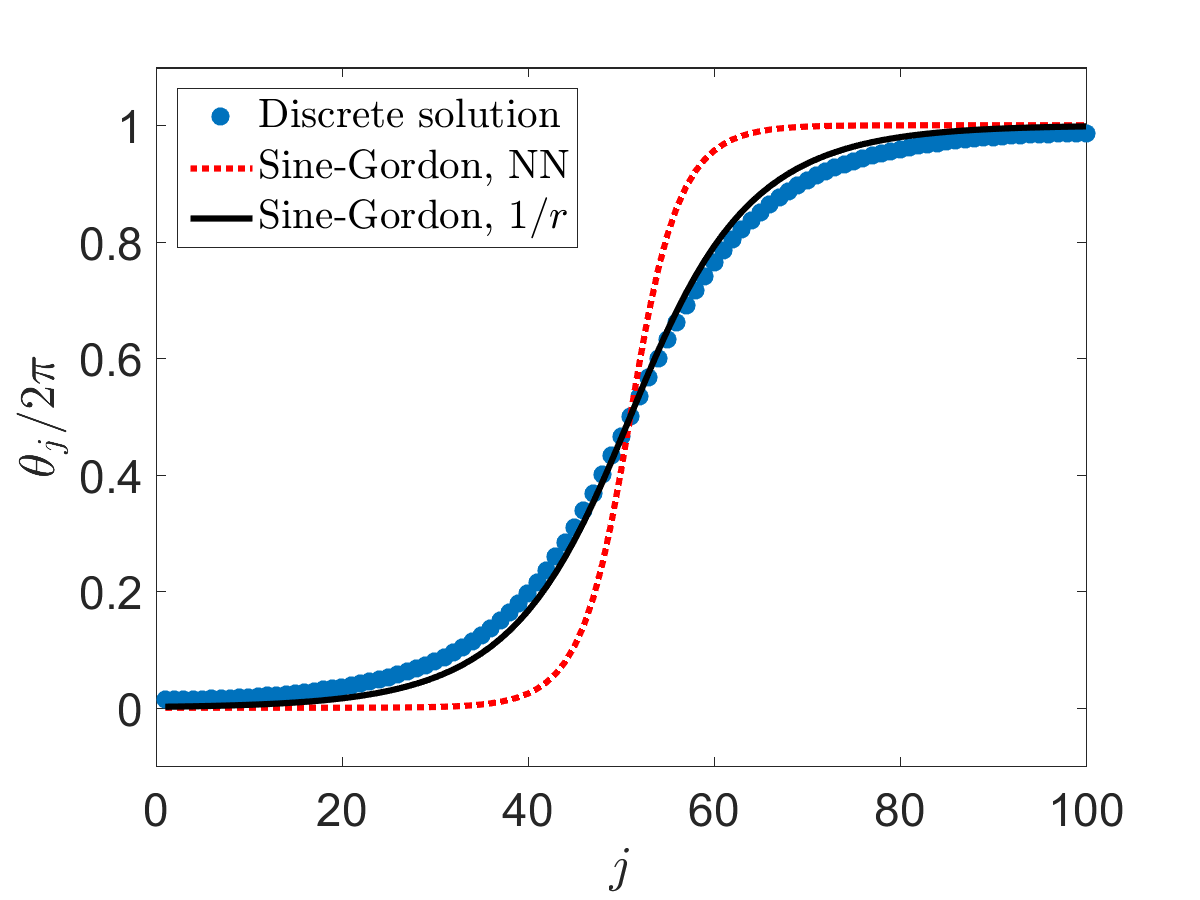}}
\subfigure[\,$N=301$]{\includegraphics[trim = 0cm 0cm 0.cm 0cm, clip,width=0.32\textwidth]{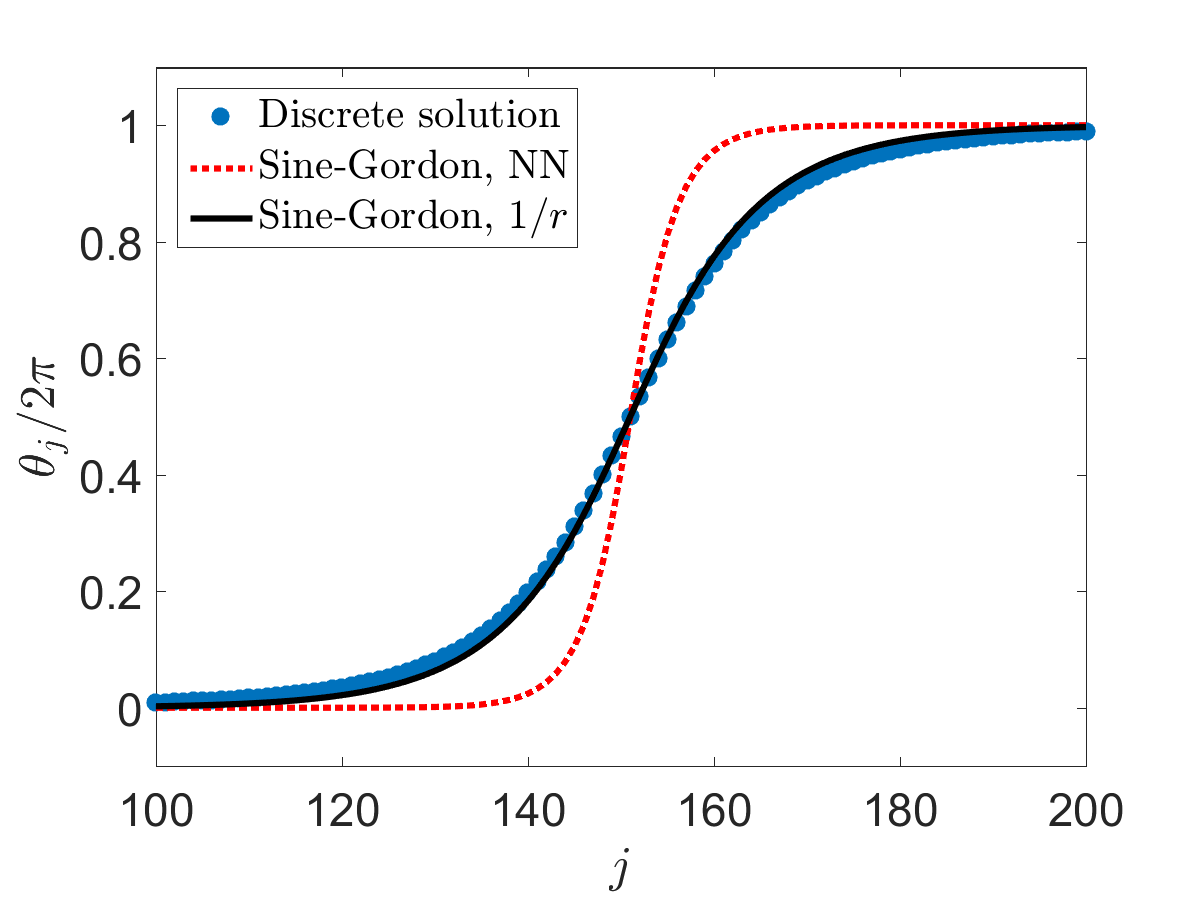}}
\bigskip
\subfigure[\,$N=1001$]{\includegraphics[trim = 0cm 0cm 0.cm 0cm, clip,width=0.32\textwidth]{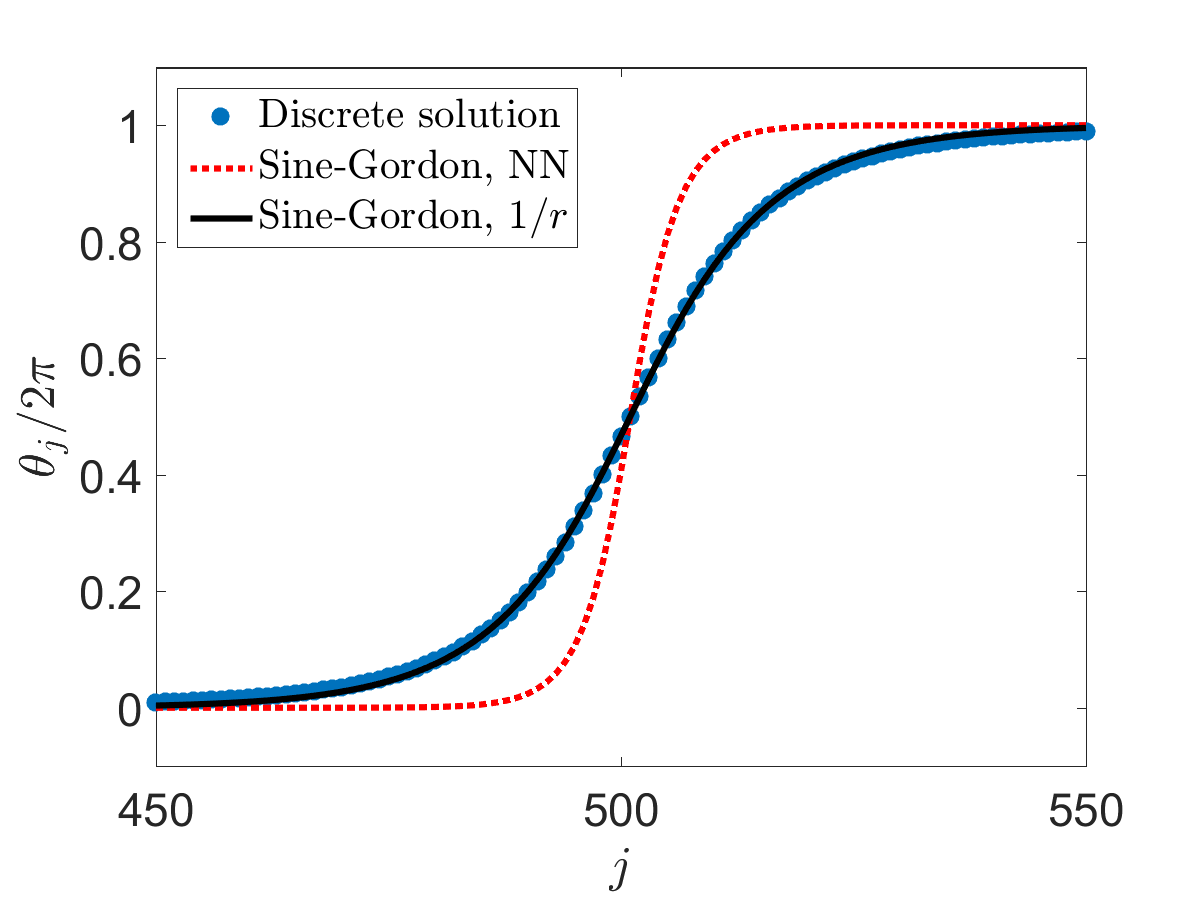}}
\caption{The normalized phase variable $\theta_j/2\pi$ of particle $j$ [Eq.~\eqref{eq:thetadef0}], measuring the particle's deviation in the kink solution from its ground state position along the chain. The circles are the result of numerically finding an equilibrium solution of Eq.~\eqref{eq:kink:discrete}, with $K_n=12,n=1,m=1,V_0=1,a=2\pi,b=2\pi$. The dotted red line is obtained by plotting the analytic sine-Gordon kink with nearest-neighbor (NN) coupling (with the kink width $d$ given in Eq.~\eqref{eq:d} taking $K_\infty=K_1=12$). The dashed black line is obtained using the same analytic expression with the kink width $\bar{d}_1\propto({\frac{3}{2}+\log N})^{1/2}$ in Eq.~\eqref{eq:bar_d_1}, setting $\alpha=0$. The number of simulated particles in the numerical solution is increased for successive panels with (a) 101, (b) 301, and (c) 1001 particles, with open boundary conditions. In (b) and (c), only the 100 particle at the center are plotted. The convergence of the numerical solution toward the analytic formula is visible.} \label{Fig:Numerics}
\end{figure}
\end{widetext}

The integro-differential equations \eqref{Kink:n} and \eqref{Kink:LR} are characterised by a left-hand side (LHS), which is a SG equation, and a the right-hand side (RHS), which is an integral term depending on the kink.
We first observe that a priori the integral term on the RHS cannot be discarded. This term, in particular, is responsible for the behaviour at the edges of the chain, far away from the kink's core: Simple considerations show that the tail of the kink decays algebraically with $1/x^{n+1}$.
This behaviour is in contrast to the exponential decay of the sine-Gordon kink at the asymptotics in Eq.~\eqref{eq:thetaSG}. It is recovered by inspecting the behaviour of the integral term at distances much larger than the kink core, where one can replace the kink's first derivative with a Dirac-delta function \cite{Braun:1990}, or can be derived from the general properties of the interactions \cite{Pokrovsky:LR}. 
In this limit, in particular, the second derivative on the LHS can be neglected. Thus, the integral term is majorly responsible for the non-local properties of the kink and gives rise to a power-law interactions between distant kinks \cite{Pokrovsky:LR,Braun:1990}.

At the kink's core the integral term is negligible for $n > 1$ as long as the ratio $a/d$ is sufficiently small. Analytical
estimates and numerical calculations indicate that the RHS of Eq. \eqref{Kink:n} scales approximately like $a/d$ for $n=2$ and has a sharper decay
for larger $n$. Therefore, for $n>1$ the integral term scales like discreteness effects at the kink's core, and in the continuum limit
the kink's core is determined to good approximation by a sine-Gordon equation with a rescaled kink's width 
\begin{equation}
\bar d_{n>1}=\frac{d_n}{\sqrt{n-1}}=\frac{a}{\sqrt{n-1}}\sqrt{\frac{K_n}{V_0}\left(\frac{b}{2\pi}\right)^2}\,.
\end{equation}
It is also interesting to analyse the scaling of $d_n$ with $n$: For $n$ finite it increases as $n$ decreases: In fact, the repulsive 
interactions become locally increasingly strong. The nearest-neighbour case can be recovered by appropriately rescaling the elastic constant
$K_n$, such as $K_n\sim n K_\infty$. In this limit the term on the right-hand side of Eq. \eqref{Kink:n} tends to 
zero and one recovers the sine-Gordon equation.

The behaviour at the kink's core for $n=1$, corresponding to repulsive Coulomb interactions, shall be discussed apart. For this case we have considered a SG kink and numerically verified that this slowly converges to the solution of Eq. \eqref{Kink:LR} as $a/d$ becomes smaller. 
For finite but large chain the integral term can be approximated by the function $-\alpha \sin \theta$, plus a correction which is negligible at the kink's core. The kink's equation at the core is then given by a SG equation, with the kink's width 
\begin{equation}
\bar d_{n=1}=\frac{a}{\sqrt{1+\alpha}}\sqrt{\frac{3}{2}+\log N}\sqrt{\frac{K_1}{V_0}\left(\frac{b}{2\pi}\right)^2}\,.\label{eq:bar_d_1}
\end{equation}
where $\alpha=\alpha(a/d)$ and $1>\alpha>0$. This coefficient monotonously decreases with $a/d$ for the values we checked. These predictions are in excellent agreement with the numerical results with a discrete chain of ions. Figure \ref{Fig:Numerics} displays the kink's solution for $N=100, 300, 500$. The solid blue line corresponds to a Sine-Gordon kink whose width is given by Eq. \eqref{eq:bar_d_1}, the convergence of the behaviour at the kink's center with the SG kink is visible. On the basis of these numerical analysis we conclude that the kink at the core is described by a SG kink whose width is proportional to $\sqrt{\log N}$ and thus weakly depends on the chain's length. 

This dependence of the kink's width on the ions number is a consequence of the weak non-additivity of Coulomb systems in one dimension. Extensivity can be formally re-established for instance, by rescaling the spring constant as
\begin{equation}
\label{Kac}K_1=K/\log N\,,
\end{equation} 
which is an implementation of Kac's scaling for one-dimensional Coulomb interactions \cite{Kac,Morigi:2004:PRL}. For a static kink, this is equivalent to increasing the depth of the 
potential as $V_0\to V_0\log N$. This rescaling leads to the definition:
\begin{equation}
\label{d:Kac}
\bar d_1^{\rm Kac}=a\sqrt{\frac{K}{V_0}\left(\frac{b}{2\pi}\right)^2}\,.
\end{equation}
Thus, in the thermodynamic limit the core of the static kink is described by a SG kink with width \eqref{d:Kac}, while at the tails the kink decays as $1/x^2$. 

\section{Conclusions}
\label{Sec:4}

We have shown a procedure that permits to determine the equation for a static kink in a chain of 
atoms interacting with a repulsive, power-law potential, and to infer the properties of the solution. Starting from the discrete equation we 
have taken a continuum limit and cast the equation into the sum of a local term, which has the form of a SG equation, and a integral term.
We have argued that in the continuum limit the SG equation determines the properties at the kink's core, while the integral gives the behaviour at the tails.
The correction of the integral term to the behaviour at the kink's core, in particular, are of the same order of the discreteness effects. These effects modify the kink's 
width and form. Moreover, they significantly modify the dynamical properties \cite{Willis,Vladan:preprint}. 

The formalism discussed here can be extended by including a non-homogeneous density distribution, 
as it is the case in the presence of a harmonic trap. In this case the gradient expansion shall be 
performed including the density in the continuum limit  
and will generally give rise to first-order derivatives of the phase function \cite{Morigi:2004}. 

The experimental study of our findings could be pursued with several physical systems. With trapped ions, for instance, recent years saw significant advances in the trapping of long chains \cite{PhysRevA.95.013413}, and subjecting trapped ions to optical lattices (though still with smaller numbers of ions \cite{PhysRevA.99.031401}). These findings could also be observed in dipolar gases \cite{Ferlaino} and Rydberg excitons \cite{Rydberg:Review}  and in the future perhaps also in magnetically repelling colloids \cite{straube2011pattern}.

\acknowledgments
The authors are deeply grateful for the privilege they had to know and to work with Shmuel Fishman. Shmuel Fishman 
was a great scientist, a person of high moral standards, and a generous friend. This work is 
dedicated to his memory. 

We thank Eugene Demler, Thomas Fogarty, David Mukamel, and Vladimir Stojanovic, for stimulating discussions and helpful comments. We acknowledge the contribution of Andreas A. Buchheit in the initial stage of this project. This project was supported by  the German Research Foundation (the priority program No. 1929 GiRyd), the European Commission (ITN ``ColOpt''), and the German Ministry of Education and Research (BMBF, QuantERA project ``NAQUAS'').  Project NAQUAS has received funding from the QuantERA ERA-NET Cofund in Quantum Technologies implemented within the European Union's Horizon 2020 program. G.M. thanks P. Eschner-Morigi for insightful comments. C.C. acknowledges funding from grant BID-PICT 2015-2236. H.L. thanks Roni Geffen for fruitful discussions, and acknowledges support by IRS-IQUPS of Universit\'e Paris-Saclay and by LabEx PALM under grant number ANR-10-LABX-0039-PALM.

\end{document}